\documentclass[12pt]{article}
\usepackage{amsmath,macros,slett,cite,slashed}
\newcommand{\e}[0]{{\rm e}^{2 \al s^2}}
\newcommand{\al}[0]{V \fpi^2 \mui^2}
\newcommand{\dl}[0]{{\rm d}s\;}
\newcommand{\mui}[0]{\mu_{iso}}
\newcommand{\lmp}[1]{\lambda_{+#1}}
\newcommand{\lmm}[1]{\lambda_{-#1}}
\newcommand{\Dp}[0]{D_+}
\newcommand{\Dm}[0]{D_-}
\newcommand{\srp}{\left \langle \sum_n \frac{1}{\lmp{n}^2+ m_u^2}\right \rangle_\nu}
\newcommand{\srm}{\left \langle \sum_n \frac{1}{\lmm{n}^2+ m_d^2}\right \rangle_\nu}
\newcommand{\xu} {x_u}
\newcommand{\xd} {x_d}
\newcommand{\ssect}[1]{\noindent {\bf  #1\ }}

\begin{document}
\begin{titlepage}
%preprint numbers
\begin{flushright}
%PREPRINT NUMBERS
\end{flushright}
%
%title
\vskip 1.2cm
\begin{center}
{\Large\bf 
Determining $\fpi$ from spectral sum rules
\\[0.5ex]
}
\end{center}
%
%authors
\vskip 1.7cm
\begin{center}
{ \large
Magdalena Luz$^{\scriptscriptstyle a}$
}
\vskip 0.5cm
%
%Institution
{%\small
%$^{\scriptstyle a}$ 
Niels Bohr Institute\\
Blegdamsvej 17\\ DK-2100 Copenhagen \O\\
Denmark
\vskip 1.0ex
}
%
%abstract
\vskip 1.575cm
{\bf Abstract}
\vskip 0.1ex
\end{center}
{\small
We derive spectral sum rules for a system with two quarks coupled to an 
imaginary isospin chemical potential in the $\epsilon$ regime. The sum rules show an explicit dependence on
the pion decay constant which should make it possible to measure $\fpi$ from the eigenvalue spectrum of this 
particular Dirac operator.
}
%%keywords
%\vskip 4.0ex
%\noindent{\it Key words:} 

%%PACS numbers
%\vskip 2.0ex
%\noindent{\it PACS:} 

\vskip 0.29cm
\vfill

%date Month year
\begin{center}
July 2006
\end{center}
\vskip 0.9cm
 $^{\scriptscriptstyle a}$ 
{\footnotesize Also at the Niels Bohr International Accademy, Blegdamsvej 17, DK-2100 Copenhagen, Denmark}

\eject
\vfill
\eject

\end{titlepage}

%\ssect{INTRODUCTION}
\ssect{1.}%intro
The determination of low energy constants of QCD such as the pion decay constant $\fpi$ remains an important problem.
In particular the computation of such quantities from lattice calculations is notoriously difficult due to the exceeding
computational challenge posed by simulations of small quark masses. In addition when approaching the chiral limit on the lattice, 
finite size effects inevitably become more and more significant.
Hence, a method which takes finite size scaling explicitly into account seems to be highly profitable.
Such a technique is provided by the so called $\epsilon$-regime
of QCD \cite{Gasser:1987ah}. This regime applies in a region where the Compton wave length of the pion $1/m_\pi$ is larger 
than the one dimensional size $L$ of the physical volume $V = L^4$, while still being much smaller than 
the typical hadronic scale 
$\Lambda_{QCD}$, i.e. $1/m_\pi > L \gg 1/\Lambda_{QCD}$. 
The lowest order %(in the particular power counting of the $\epsilon$-regime)
effective partition function of the $\epsilon$-regime is known analytically.
The fact that
it depends explicitly on the infinite volume 
chiral condensate $\Sigma= \langle\psibar \psi \rangle$ can be exploited to determine this constant 
with high accuracy:
%has drawn much attention to this regime of QCD and a  
%variety of methods %\cite{} 
%were developed to extract that quantity from calculations in the $\epsilon$-regime. 
back in 1992 Leutwyler and Smilga derived a set of spectral sum rules for the Dirac operator  
by restricting the partition function to sectors with fixed topological charge $\nu$ \cite{ls}.
In this way the chiral condensate is linked to the spectrum of the
Dirac operator in finite volume which can be determined from lattice simulations \cite{Shuryak, Ogawa}.

However, with the standard Dirac operator other low energy constants such as the pion decay constant 
$\fpi$ appear only in higher order corrections \cite{Damgaard:2001js}.  
%of the effective partition function. 
As a consequence, the computation of $\fpi$ was believed to be much more demanding 
%and the need for very high statistics made them much more demanding 
\cite{Bietenholz, Giusti, Ogawa}. 
It is therefore still more widespread to extrapolate the pion decay constant down to the chiral
limit from simulations in the $p$ regime 
of chiral perturbation theory \cite{fpi_q, fpi_uq}.

Recently a new approach has been proposed 
\cite{heller,Damgaard:2006pu,mehen} which avoids these difficulties: 
if the quarks are coupled to an external source,
dependency on $\fpi$ appears already at the lowest order in the partition function
\cite{verbaarschot,Damgaard:2001js}.
The authors of Refs.~\cite{heller,Damgaard:2006pu} have used this fact to derive correlation functions of the eigenvalue 
densities which are sensitive to $\fpi$ for both quenched and unquenched chiral perturbation theory. 
Here, we will use the same partition function to deduce a set of sum rules in a way 
analogous to Ref.~\cite{ls}. These rules depend then likewise on $\fpi$ and make it possible to determine this important 
quantity from a lattice simulations in finite volume.

\ssect{2.}
We consider a system with two quark flavours $u$ and $d$ which are coupled to an external source $\mui$. The source 
can be interpreted as an imaginary isospin chemical 
potential which couples differently to the two flavours 
or as twisted boundary conditions \cite{mehen, Kogut}. 
This gives rise to two independent eigenvalue equations 
\bea
(\Dp + m_u) \; \psi_{+n} = (\slashed{D}[A] + i \mui \gamma_0 + m_u)\; \psi_{+n} &= (i\lmp{n} + m_u)\; \psi_{+n}\\
(\Dm + m_d) \; \psi_{-n} = (\slashed{D}[A] - i \mui \gamma_0 + m_d)\; \psi_{-n} &= (i\lmm{n} + m_d)\; \psi_{-n},
\eea
where $A$ denotes the gauge field. 
The advantage of an {\it imaginary isospin} chemical potential is twofold. Firstly, an isospin chemical 
potential preserves the positivity of the the fermion matrix \cite{Kogut}.
The system can thus be simulated on the lattice with the usual Monte Carlo techniques without running into 
sign problems. 
Secondly, if it is chosen to be imaginary the massless operators $\Dp$ and $\Dm$ are anti-hermitian and 
the eigenvalues $\lambda_\pm$ lie on the real axis. 

In the case of degenerate masses $m_u = m_d$ the two flavours can be converted into each other by 
the transformation
$\mui$ to $-\mui$.
Here we use a non mass-degenerate quark pair for purely technical reasons: 
as we will see below this gives us the only handle to distinguish between the two sets of eigenvalues.
Thus keeping the masses distinct allows us to 
extract sum rules for only one set of eigenvalues, $\lmp{}$ or $\lmm{}$ accordingly. 

For both sets of eigenvalues %$\lmp$ and $\lmm$ 
the non zero modes come in positive and negative pairs, $\pm \lmp{n}$ and $\pm \lmm{n}$ respectively.
Hence in a sector of fixed topology,
the partition function can be written as
%$\lmp{}$ and $\lmm{}$
\begin{equation}
\label{zge}
Z_\nu(m_u, m_d) = \left \langle m_u^\nu \; m_d^\nu\;  \prod_n(\lmp{n}^2 + m_u^2)\;\prod \limits_{n} (\lmm{n}^2 + m_d^2)\right \rangle_\nu,
\end{equation}
where $\langle \cdot \rangle_\nu$ denotes the gauge average over all configurations with topological charge $\nu$ and the products 
are restricted to strictly positive values of $\lambda$.
 
Our derivation of the sum rules deviates from
the procedure originally used by 
Leutwyler and Smilga in Ref.~\cite{ls}, we will
follow the somewhat easier approach presented in Ref.~\cite{dmg} instead. 
%The sum rules of Ref.~\cite{ls} were derived by expanding this formula for small masses. Here, we will take a slightly 
%different approach which does not depend on the masses being small:
Massive spectral sum rules for $\lmp{}$ (or $\lmm{}$) can be derived from formula (\ref{zge}) by taking logarithmic derivatives 
with respect to the
masses $m_u$ (or $m_d$) 
%\cite{dmg}
. A first order sum rule for instance is given by
\begin{equation}
\label{sr_1o}
\frac{1}{2m_u} \left(\frac{\partial}{\partial m_u} \ln Z_\nu (m_u, m_d) - \frac{\nu}{m_u}\right) =\; \srp.
\end{equation}
%(or the analogous expression for $\lmm{n}$).
As in the products of Eq.~(\ref{zge}) the sum on the rhs runs over the positive eigenvalues only.

On the other hand 
in the $\epsilon$ expansion of chiral perturbation theory the partition function can be
shown \cite{verbaarschot, heller} to be
\begin{equation}
\label{zex}
Z_\nu (\xu, \xd) = {\rm e}^{-2\al} \int_0^1 \dl \e s\; I_\nu(s \xd) I_\nu(s \xu),
\end{equation}
where we introduced the scaling variables of the $\epsilon$-regime $x_i = V \Sigma m_i,\;  i = u, d$ and
$I_\nu$ are modified Bessel functions.
The dependence on $\fpi$ is through the product $V \fpi^2 \mui^2 $ only.
In particular a sign change in $\mui$ leaves the partition function invariant. The two flavours are indeed only distinguishable
through their masses.
Upon inserting this explicit formula for $Z_\nu$ into Eq.~(\ref{sr_1o}), we obtain the first order sum rule
\be
\label{sr_o1_ex}
\begin{split}
\srp  = \frac{V^2\Sigma^2}{2 \xu}\frac{\int_0^1 \dl \e \; s^2  \;   I_\nu(s \xd)\;  I_{\nu+1} (s  \xu)}
{\int_0^1 \dl \e \; s \; I_\nu(s \xd)\; I_\nu(s \xu)}.
\end{split}
\ee
This formula simplifies a good deal if we take both quark masses to zero. The Bessel functions disappear up to remnant
powers of the parameter $s$ 
\be
\label{sr_o1_ml} 
%\begin{split}
\left \langle \sum_n \frac{1}{\lambda^{2}_{+n}}\right \rangle_\nu 
= \frac{V^2 \Sigma^2}{4(\nu+1)}\frac{\int_0^1 \dl \e s^{2\nu + 3} }{\int_0^1 \dl \e s^{2\nu+1}} .
%&= \frac{V \Sigma^2}{8\fpi^2\mui^2(\nu+1)}
\ee
For general $\nu$ this expression can only
 be evaluated in terms of incomplete and ordinary $\Gamma$-functions 
\be
\label{sr_o1_g}
\begin{split}
\left \langle \sum_n \frac{1}{\lambda^{2}_{+n}}\right \rangle_\nu 
&=-\frac{V \Sigma^2}{8(\nu +1) \fpi^2 \mui^2}\frac{\Gamma(\nu+2) - \Gamma(\nu+2, -2\al)}{\Gamma(\nu+1) - \Gamma(\nu+1,-2\al) }=\\
&=-\frac{V\Sigma^2}{8 (\nu + 1)\fpi^2\mui^2}\frac{\int_{-2\al}^0 t^{\nu+1} {\rm e}^{-t} dt}{\int_{-2\al}^0 t^{\nu} {\rm e}^{-t} dt}.
\end{split}
\ee
However, for a given topological charge it is straightforward to calculate the parameter integrals. The rule reduces then to a 
rather simple expression, where $\fpi$ appears only in polynomials and exponentials and which can easily be fitted to
lattice data. 
Let us illustrate this for the case of vanishing topological charge, there we simply have
%For instance in the sector of vanishing topological Eq.~(\ref{sr_o1_ml}) simply is 
%as an illustration
\be
\left \langle \sum_n \frac{1}{\lambda^{2}_{+n}}\right \rangle_0 = \frac{V \Sigma^2}{8  \fpi^2 \mui^2 }\frac{1 + {\rm e}^{2 \al} (2\al-1)}{{\rm e}^{2\al}-1}.
\ee
Eqs.~(\ref{sr_o1_ml}) and~(\ref{sr_o1_g}) are the direct equivalents of the first order sum rule given in 
Ref.~\cite{ls}, but evaluated 
for a system with two 
quarks coupled the chemical potential $\mui$. Indeed, in the limit where $\mui$ vanishes
Eq.~(\ref{sr_o1_ml}) becomes
\begin{equation}
\left \langle \sum_n \frac{1}{\lmp{n}^{2}}\right \rangle_\nu 
= \frac{V^2 \Sigma^2}{4(\nu+1)}\frac{\int_0^1 \dl s^{2\nu + 3} }{\int_0^1 \dl s^{2\nu+1}} = \frac{V^2 \Sigma^2}{4(\nu + 2)},
\end{equation}
which is precisely the Leutwyler-Smilga result for $N_f =2$ flavours.

Due to the symmetry of the partition function under an exchange of the quark masses, 
the corresponding sum rule for $\lmm{}$ can be obtained from Eq.~(\ref{sr_o1_ex}) by simply substituting $x_u$ for $x_d$.
If we had treated the quarks as mass degenerate from the beginning we would have arrived at the sum of those
two sum rules, which is
just twice Eq.~(\ref{sr_o1_ex}).

\subsubsection*{Higher order sum rules}
We can derive two different types of second order sum rules by either taking the second derivative with respect to $m_u$,
$\frac{\partial^2}{\partial m_u^2} \ln Z_\nu$, or a mixed derivative $\frac{\partial^2}{\partial m_u \partial m_d} \ln Z_\nu$.
Let us start with the second possibility. Applied to Eq.~(\ref{zge}) the mixed derivative 
yields the subtracted correlation of
the two sets of eigenvalues 
\be 
\label{sr_o2_m}
\begin{split}
\frac{1}{4 m_u m_d} \left[\frac{\partial^2}{\partial m_u \partial m_d} \ln Z_\nu (m_u, m_d)\right] &= 
\left\langle \left(\sum_n \frac{1}{\lmp{n}^{2} + m_u^2}\right)
\left (\sum_n \frac{1}{\lmm{n}^{2} + m_d^2}\right)\right \rangle_\nu \\&- \srp \srm
\end{split}
\ee
%\be
%\label{sr_o2_ex}
%\begin{split}
%&\frac{1}{4 \; m_u\;  m_d} \left[\frac{\partial^2}{\partial m_u \partial m_d} \ln Z_\nu (m_u, m_d)\right] =\\ 
%&=\frac{V^4 \Sigma^4}{4 \xu \xd} \left[\frac{\int_0^1 dl \e l^3 I_{\nu + 1} (l\xu) I_{\nu + 1} (l\xd)}{\int_0^1 dl \e l I_\nu(l \xu) I_\nu (l \xd)} \right. \\ 
%&-\left.\frac{\int_0^1 \dl l^2  \; \e \;  I_\nu(l\; \xd)\;  I_{\nu+1} (l \; \xu)\int_0^1 \dl l^2  \; \e \;  I_{\nu+1}(l\; \xd)\;  I_{\nu} (l \; \xu)}{[\int_0^1 \dl l \; \e \; I_\nu(l\; \xd)\; I_\nu(l\; \xu)]^2} 
%\right]
%\end{split}
%\ee
 Since we have calculated the disconnected parts already, the only new contribution from Eq.~(\ref{sr_o2_m}) is a sum rule for
\be
\begin{split}
\left\langle \left(\sum_n \frac{1}{\lmp{n}^{2} + m_u^2}\right)\right.&\left.
\left (\sum_n \frac{1}{\lmm{n}^{2} + m_d^2}\right)\right \rangle_\nu =\\
&=\frac{V^4 \Sigma^4}{4 \; \xu \xd} \frac{\int_0^1 \dl \e s^3\; I_{\nu + 1} (s\xu) I_{\nu + 1} (s\xd)}
{\int_0^1 \dl \e s\;  I_\nu(s \xu) I_\nu (s \xd)}.
\end{split}
\ee
In the limit of vanishing quark masses we can again expand the Bessel functions and obtain a much simpler expression
for the massless mixed second order sum rule
\be
\label{sr_o2_m_ml}
\begin{split}
\left\langle \left(\sum_n \frac{1}{\lmp{n}^{2}}\right)
\left (\sum_n \frac{1}{\lmm{n}^{2}}\right)\right \rangle_\nu&=
\frac{V^4 \Sigma^4}{16(\nu + 1)^2} \frac{\int_0^1 \dl \e \; s^{2\nu + 5}}{\int_0^1 \dl \e \; s^{2\nu + 1}}=\\
&= \frac{V^2 \Sigma^4}{64\fpi^4\mui^4(\nu+1)^2}
\frac{\Gamma(\nu+3) - \Gamma(\nu+3, -2\al)}{\Gamma(\nu+1) - \Gamma(\nu+1,-2\al) }.%\\
%&=\frac{V^2\Sigma^4}{64\fpi^4\mui^4(\nu+1)^2}\frac{\int_{-2\al}^0 t^{\nu+2} {\rm e}^{-t} dt}{\int_{-2\al}^0 t^{\nu} {\rm e}^{-t} dt}.
\end{split}
\ee
Apart from the coefficients, the difference to the massless first order rule is given by the higher power
of the parameter $s$ in the first line of Eq.~(\ref{sr_o2_m_ml}). These powers are indeed a distinctive feature for any massless sum rule 
of a given order (third order sum rules for instance carry a power of $s^{2\nu+7}$).
Concerning the evaluation of the integral %in Formula~(\ref{sr_o2_m_ml}) 
we can make exactly the same remarks 
as for the first order rule, they are easily calculated in a fixed topological sector.
For completeness we give again the result at $\nu = 0$
\be
\left\langle \left(\sum_n \frac{1}{\lmp{n}^{2}}\right)
\left (\sum_n \frac{1}{\lmm{n}^{2}}\right)\right \rangle_0=
\frac{V^2 \Sigma^4}{32 \fpi^4 \mui^4}\frac{(1 -  2\al(1 +  V \fpi^2 \mui^2)){\rm e}^{2\al} - 1}{{\rm e}^{2 \al} -1} .
\ee 
Again, in the limit $\mui \to 0$ where $\lmp{}$ and $\lmm{}$ become degenerate the formula reproduces the result of Ref.~\cite{ls} 
In addition, this second order sum rule can be compared to the results of Ref.~\cite{heller}. There 
the mixed two point spectral correlation function
\be 
\begin{split}
\rho^{(2)}(\lambda_1, &\lambda_2, m_u, m_d, i\mui) = \left \langle \sum_n \delta(\lambda_1 - \lmp{n})\sum_l \delta(\lambda_2 -\lmm{l})
\right \rangle \\
&- \left \langle \sum_n \delta(\lambda_1 - \lmp{n})\right \rangle \left \langle \sum_l \delta(\lambda_2 -\lmp{n})\right\rangle.
\end{split}
\ee
is derived with the replica method.
By integrating out $\lambda_1$ and $\lambda_2$ in 
\be
\label{rhox2y2}
\begin{split}
\int d\lambda_1 \int d\lambda_2 &\; \frac{\rho^{(2)}(\lambda_1, \lambda_2, m_u,m_d, i\mui) }{\lambda_1^2 \lambda_2^2} 
%= \frac{1}{4 m_u m_d} \left[\frac{\partial^2}{\partial m_u \partial m_d} \ln Z_\nu (m_u, m_d)\right].
%&= 
%\left\langle \left(\sum_n \frac{1}{\lmp{n}^{2} + m_u^2}\right)
%\left (\sum_n \frac{1}{\lmm{n}^{2} + m_d^2}\right)\right \rangle_\nu - \srp \srm
\end{split}
\ee
we should reproduce our subtracted mixed sum rule of Eq.~(\ref{sr_o2_m}). 
Indeed, we do find exact agreement in the massless limit.
%where the density is given by
%with the density 
%$\rho^{(2)}$ 
%
%massless rho
%\be
%\begin{split}
%&\rho^{(2)}(\lambda_1, \lambda_2, m_u, m_d, i\mui)|_{m_{u/d}=0} = \lambda_1  \lambda_2 \left[\int_0^1 \dl  \e s^{2\nu+1}\right]^{-2} \\
%&\cdot \left[\int_0^1 \dl  \e s J_\nu(s \lambda_1)J_\nu(s \lambda_2) \int_0^1 \dl \e s^{2\nu+1} \right.\\
%&- \left.\int_0^1 \e s^{\nu+1} J_\nu(s\lambda_1)\int_0^1 \dl \e s^{\nu+1} J_\nu(s\lambda_2)\right]\cdot\\
%&\cdot\left[\int_0^1 \dl \e s^{2\nu + 1}\int_1^{\infty} \dl  {\rm e}^{-2\al s^2} s J_\nu(s\lambda_1)J_\nu(s\lambda_2) + 
%\frac{J_{\nu+1}(\lambda_1)J_{\nu+1}(\lambda_2)}{\lambda_1 \lambda_2}\right]. 
%\end{split}
%\ee

With the first choice above $\frac{\partial^2}{\partial m_u^2} \ln Z_\nu$, we have means to extract the correlation 
between eigenvalues of one series $\lmp{}$ (or equivalently $\lmm{}$).
More precisely, from taking the second derivative with respect to $m_u$ and subtracting the zero mode contributions and the 
known first order terms, we can deduce a formula for
\begin{equation}
\label{sr_o2_p}
\begin{split}
 &\left\langle  \left(\sum_n \frac{1}{\lmp{n}^{2} + m_u^2}\right)^2 
- \sum_n \left(\frac{1}{\lmp{n}^{2} + m_u^2}\right)^2 \right\rangle_\nu =\\
&=  \frac{V^4 \Sigma^4}{4\; \xu^3} \frac{\int_0^1 \dl \e \; s^2 \;I_\nu(s \xd) \left[ \; s \; \xu \; I_\nu(s\xu) - 
2(\nu + 1)I_{\nu + 1} (s \xu)\right]}{\int_0^1 \dl \e \; s\;  I_\nu(s \xd) I_\nu(s\xu)}. 
\end{split}
\end{equation} 
Again the formula simplifies drastically in the massless limit where we get
\begin{equation}
\label{sr_o2_p_ml}
\begin{split}
\left\langle  \left(\sum_n \frac{1}{\lmp{n}^{2}}\right)^2 
-\right.&\left. \sum_n \frac{1}{\lambda_{+n}^{4}} \right\rangle_\nu 
=\frac{V^4 \Sigma^4}{16 (\nu + 1) (\nu + 2)}\frac{\int_0^1 \dl \e \; s^{2\nu + 5}}{\int_0^1 \dl \e \; s^{2\nu + 1}}\\
%&= \frac{V^2 \Sigma^4}{64\fpi^4\mui^4(\nu+1)(\nu+2)}
%\frac{\Gamma(\nu+3) - \Gamma(\nu+3, -2\al)}{\Gamma(\nu+1) - \Gamma(\nu+1,-2\al) }\\
%&=\frac{V^2\Sigma^4}{64\fpi^4\mui^4(\nu+1)(\nu+2)}\frac{\int_{-2\al}^0 t^{\nu+2} {\rm e}^{-t} dt}{\int_{-2\al}^0 t^{\nu} {\rm e}^{-t} dt}.
\end{split}
\end{equation}
Let us remark here that at finite $\mui$ we cannot  
extract a separate sum rule for the fourth order term 
$\langle \sum 1/\lmp{}^4\rangle_\nu$ from the partition function in Eq.~(\ref{zex}).
Since the $u$ and the $d$ quark obey two different eigenvalue equations, each of them behaves effectively like a
one flavour system.
In order to isolate the fourth order term above we would need another flavour with the same eigenvalues $\lmp{n}$
but different mass $\tilde{m}_u$. At vanishing 
isospin chemical potential however, where $\lmp{}$ and $\lmm{}$ become degenerate
a fourth order term can be calculated from the difference of the two
second order sum rules described here and it again coincides with the result of Ref.~\cite{ls}. 

It is in principle possible to continue this series and to calculate sum rules of any order 
by taking higher and higher derivatives, and subtracting
the known lower order terms. The computation however becomes tedious and its usefulness doubtful.

\ssect{3.} %\ssect{CONCLUSION}
In this paper we have derived a set of spectral sum rules for a system of two quarks coupled to 
an imaginary isospin chemical potential from the finite volume partition function.
This can be seen as an extension of the pioneering work of Leutwyler and Smilga in Ref.~\cite{ls}.  
The sum rules derived here inherit the $\fpi$ dependence from the partition function and provide means to
determine this important constant from the spectrum of the Dirac operator introduced here.

%The dependence of the finite volume partition function on $\fpi$ readily 
\subsubsection*{Acknowledgements} The author thanks P.H. Damgaard for useful discussions and a careful reading of the manuscript.
Part of this work was accomplished during the Doctoral Training Programme 2006 at ECT* in Trento. Financial support 
from ECT* is gratefully acknowledged.

\end{document}